\documentclass[twocolumn,showpacs,superscriptaddress,prl]{revtex4}
\usepackage{graphicx}%
\usepackage{dcolumn}
\usepackage{amsmath}

\makeatletter
\def\btt#1{\texttt{\@backslashchar#1}}%
\DeclareRobustCommand\bblash{\btt{\@backslashchar}}%
\makeatother

\begin{document}

\preprint{HEP/123-qed}

\title{Measurement of the Cosmic-Ray Antiproton to Proton Abundance Ratio\\
between 4 and 50 GeV}

\author{A.S.~Beach}
\author{J.J.~Beatty}
\affiliation{Department of Physics,
Pennsylvania State University, University Park, PA 16802}
\author{A.~Bhattacharyya}
\affiliation{
Department of Physics, Indiana University, Bloomington, IN 47405}
\author{C.~Bower}
\affiliation{
Department of Physics, Indiana University, Bloomington, IN 47405}
\author{S.~Coutu}
\affiliation{Department of Physics,
Pennsylvania State University, University Park, PA 16802}
\author{M.A.~DuVernois}
\thanks{Now at School of Physics and Astronomy,
University of Minnesota, MN 55455}
\affiliation{Department of Physics,
Pennsylvania State University, University Park, PA 16802}
\author{A.W.~Labrador}
\thanks{Now at California Institute of 
Technology, Pasadena, CA 91125}
\affiliation{Enrico Fermi Institute and Department of Physics,
University of Chicago, Chicago, IL 60637}
\author{S.~McKee}
\affiliation{
Department of Physics, University of Michigan, Ann Arbor, MI 48109}
\author{S.A. Minnick}
\affiliation{Department of Physics,
Pennsylvania State University, University Park, PA 16802}
\author{D.~M\"uller}
\affiliation{Enrico Fermi Institute and Department of Physics,
University of Chicago, Chicago, IL 60637}
\author{J.~Musser}
\affiliation{
Department of Physics, Indiana University, Bloomington, IN 47405}
\author{S.~Nutter}
\thanks{Now at Physics and Geology Department,
Northern Kentucky University, Highland Heights, KY 41099}
\affiliation{Department of Physics,
Pennsylvania State University, University Park, PA 16802}
\author{M.~Schubnell}
\thanks{corresponding author: schubnel@umich.edu}
\affiliation{
Department of Physics, University of Michigan, Ann Arbor, MI 48109}
\author{S.~Swordy}
\affiliation{Enrico Fermi Institute and Department of Physics,
University of Chicago, Chicago, IL 60637}
\author{G.~Tarl\'e}

\author{A.~Tomasch}
\affiliation{
Department of Physics, University of Michigan, Ann Arbor, MI 48109}

\date{\today}

\begin{abstract}
We present a new measurement of the antiproton to proton
abundance ratio, $\overline{p}/p$, in the cosmic radiation.
The HEAT-pbar instrument, a balloon borne magnet spectrometer
with precise rigidity and multiple energy loss measurement capability,
was flown successfully in Spring 2000, at an average atmospheric
depth of 7.2 g/cm$^2$.
A total of 71 antiprotons were identified above the vertical
geomagnetic cut-off rigidity of 4.2 GV. The highest measured
proton energy was 81 GeV. We find that the
$\overline{p}/p$ abundance ratio agrees with that expected
from a purely secondary origin of antiprotons produced by
primary protons with a standard soft energy spectrum.
\end{abstract}

\pacs{95.85.Ry, 96.40.De} 
\maketitle

Antiprotons constitute a rare but interesting component of the
cosmic radiation. They are secondary cosmic-ray particles,
generated in nuclear interactions of high-energy cosmic rays
with the interstellar medium (ISM). It remains an open question
whether there are significant additional contributions that have
a different and perhaps more exotic origin. The kinematic threshold
energy for $\overline{p}$ production in p-p collisions of primary
cosmic-ray protons causes a $\overline{p}$ energy spectrum, and
a $\overline{p}/p$ intensity ratio, that decline rapidly from a
few GeV towards lower energies. Solar modulation inside the
heliosphere softens this `cutoff' and leads to uncertainties
in the predicted flux at low energy. Additional sources of antiprotons
might be evaporating primordial black holes (PBH), or annihilating
supersymmetric particles. The PBH contribution would be
expected to be significant at energies well below 1 GeV, while
supersymmetric particle annihilations,
for instance neutralinos, could also affect the
antiproton intensity at higher energy, above several GeV.
At those higher energies, the $\overline{p}$ energy spectrum will be
essentially unaffected by uncertainties due to solar modulation.

The energy spectrum of antiprotons measured near Earth
carries the imprint of losses during propagation through the Galaxy
and thus, is a sensitive probe of the confinement environment of
protons. For instance, if the propagation path length $\lambda$ and
the diffusion coefficient for protons
depended on energy E in the same way as has been observed for the
heavy cosmic ray nuclei (i.e. $\lambda \propto E^{-0.6}$),
one would expect the antiproton fraction $\overline{p}/p$  to
gradually decrease with increasing energy above a few GeV. This
behavior would be equivalent to that of the intensities of
secondary spallation
nuclei such as Li, Be, and B, relative to those of their heavier
primary parents C and O. In order to obtain a self-consistent
model of the propagation of protons in the Galaxy the observed
$\overline{p}$ spectrum must also be compared with measurements
of positrons and gamma rays, which also result from nuclear
interactions in the ISM (mostly via $\pi^+$ and $\pi^0$ decay).

It would be difficult to explain if the antiproton fraction
$\overline{p}/p$ were found to be constant or if it increased
at higher energy. Models of extragalactic origin are unlikely
because of constraints in intergalactic transport \cite{adams}. Closed
Galaxy models have been suggested \cite{peters} which would boost
the $\overline{p}/p$ ratio at high energy, but these
would also predict a higher abundance of He$^3$ than has been
observed \cite{smili,reimer}. It has been proposed that
the primary proton spectrum in distant regions of the Galaxy is
harder than locally observed \cite{msr98}. This would
also lead to enhanced  $\overline{p}$
production in these regions, although not necessarily to an
enhanced  $\overline{p}$ fraction near Earth. Thus it appears
that a constant or rising  $\overline{p}/p$ fraction might indeed require
the presence of a primary and possibly exotic source of antiprotons.

Observationally, the situation has been unclear. Measurements at
low energies, in particular the series of observations with
the BESS instrument \cite{bess00,bess_pp}, have provided
the $\overline{p}$ energy spectrum with good statistical accuracy
from $\sim$ 0.2 to 3 GeV. These results are in good agreement with
interstellar secondary production models. PBH contributions, if
they exist, are hidden
by the uncertainties due to solar modulations. Above
5 GeV, the results reported in the three previous
measurements \cite{imax,mass2,caprice94,caprice98} are
statistics limited, and no solid conclusion about the shape of the
energy dependence of the antiproton fraction can be drawn. The
current HEAT-pbar experiment has been developed to clarify this
situation with a series of balloon flights.

\begin{figure}[bp] 
\includegraphics[width=0.78\linewidth]{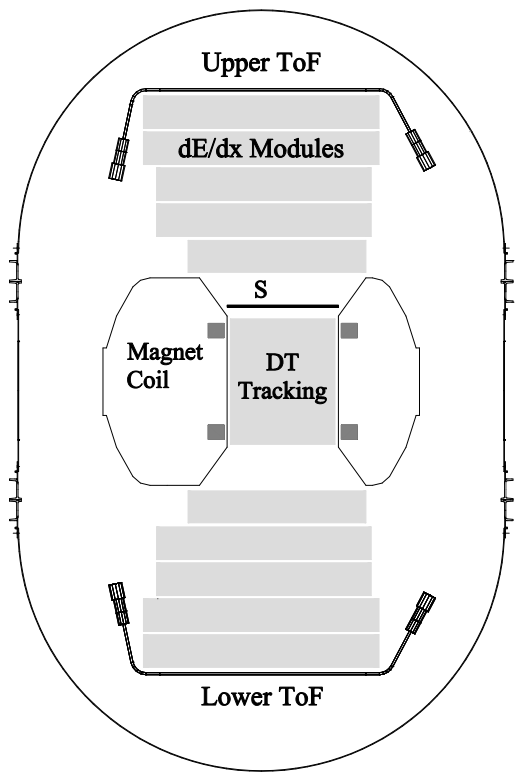}
\caption{Schematic diagram of the HEAT-pbar instrument. The
scintillator (S) above the hodoscope is part of the event
trigger. Upper and lower ToF scintillators are 2.8~m apart.}
\label{instr_fig}
\end{figure}

A schematic view of the HEAT-pbar instrument is shown in
Fig. \ref{instr_fig}. It consists of a superconducting magnet
with a drift tube hodoscope at its center, combined with stacks
of multi-wire proportional chambers above and below the hodoscope.
Two layers of scintillators, one at the very top and one at the
very bottom of the instrument, provide time-of-flight (ToF)
measurements and, together with a scintillator just above the
hodoscope, form the event trigger.

The spectrometer has been described in detail \cite{barwick97nim}.
It consists of 479 drift tubes in twenty four layers. It is
mounted in the room-temperature bore of a superconducting magnet
with a 10 kG central field, and measures the particle trajectory
within the magnetic field, providing both the particle rigidity
(momentum/charge) and charge sign. It has a single-point
tracking accuracy of better than 70 $\mu$m, an average track
length of 58 cm, and typically measures 15 points along a particle
trajectory. The continuous-tracking approximation then yields a mean
Maximum Detectable Rigidity (MDR) of 170 GV, and the rigidities
of particles up to $\sim$ 60 GV can be reliably measured. 

Antiproton flux measurements require excellent particle identification
for background discrimination. The primary sources of background to
the $\overline{p}$ flux are electrons, and negatively charged muons,
pions and kaons produced in the atmosphere as well as in the material
above and in the detector.
To provide mass discrimination between antiprotons and these particles
we measure multiple samples of the ionization energy loss. The
logarithmic rise in the mean rate of energy loss for a relativistic
charged particle is used to determine the Lorentz factor of the
particle from which, together with the rigidity measurement, the
mass is obtained.

The multiple dE/dx detector consists of a stack of 140 segmented
multi-wire proportional chambers, each providing a measurement of
the specific ionization loss. In order to maximize the particle
identification power, the proportional chambers are filled with
Xenon (and 5\% CH$_4$), which exhibits an increase in ionization
loss rate of 70\% between minimum ionization and relativistic
saturation \cite{labrador_icrc}.

The HEAT-pbar instrument was flown from Ft. Sumner, NM on June 3,
2000. The detector was at float altitude for 22 hours, at an average
atmospheric overburden of 7.2 g/cm$^2$. More than 1.9 million
events were recorded over an integrated live time of 16.2 hours at
a residual pressure between 4.5 and 8.6 mbar. The average vertical
geomagnetic cut-off rigidity along the flight path is 4.2 GV. The
instrument performed flawlessly during the flight.

Albedo particles, which mimic antiparticles in the spectrometer,
can be efficiently rejected with the ToF measurement. The flux
of upward-going relativistic particles is roughly 10$^{-3}$ that
of relativistic downward-going particles, and thus a rejection power
of 1000 is required to keep the contamination of these particles
below 1\% in the final antiproton data set. The standard deviation
in the velocity distribution for relativistic protons is 0.093c. 
This results in a rejection power against upward moving particles
which is several orders of magnitude better than required.

\begin{figure} [th]
\includegraphics [width=0.85\linewidth]{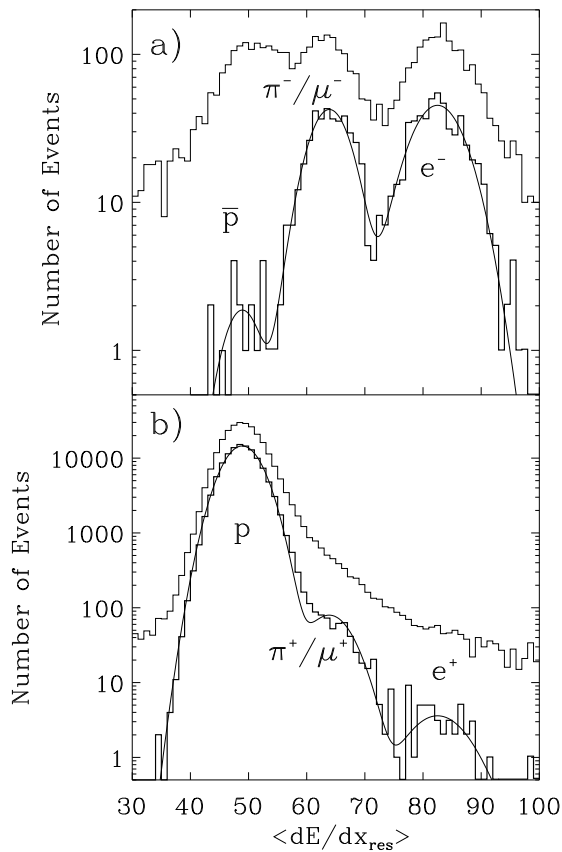} 

\caption{Histograms of the $\langle dE/dx_{res} \rangle$ response for
negative (a)
and positive (b) particles in the rigidity range 4.5 -- 6 GV. The
distributions of particle species are gaussian to
better than four order of magnitude (as can be seen
in the rising edge of the proton distribution for instance).
The upper distribution in each figure shows the corresponding
data before track quality selection criteria are applied (and
thus the majority of events in the large peak at the antiproton
position are really proton tracks out in the tails of the track
parameter distributions.)
}
\label{fig4_7}
\end{figure}

In analyzing events recorded by the dE/dx system, first
suitable selections on tracking
quality are made and events are selected for which the incident
particle transits the entire dE/dx chamber system. The effect of
high-energy tails in the Landau distribution is minimized by
computing, for each event, a restricted average specific ionization
$\langle dE/dx_{res} \rangle$ which is the average ionization signal
measured by 50\% of the dE/dx chambers with signals smaller than the
median for this event. Note that the precise value of the selected
fraction is not very critical, but that 50\% is close to optimal.
Fig.~2a and b show histograms of the $\langle dE/dx_{res} \rangle$
response for negative and positive particles, respectively,
having rigidities in the range 4.5--6 GV.
In these distributions, the peak at smallest $\langle dE/dx_{res} \rangle$
corresponds to protons and antiprotons, the next peak
to $\pi^+$/$\mu^+$ and $\pi^-$/$\mu^-$, and the peak at
large $\langle dE/dx_{res} \rangle$ to $e^+$ and $e^-$.
Compared to the antiproton flux, the kaon production is small,
although not negligible. Our Monte Carlo simulations show that our
event selection criteria reduce this contribution to
negligible levels, since kaons result from interactions in or
near the instrument.
A somewhat better mass resolution can be obtained by properly
accounting for the dependence of the $\langle dE/dx_{res} \rangle$ signal on
the rigidity within a given rigidity interval, but these
histograms, which are representative of the equivalent histograms
at higher energy, demonstrate clearly that particle identification
is achieved for both antiprotons and positrons. The large sampling
for each energy loss measurement produces highly gaussian
distributions, and thus for each energy interval, the $\overline{p}/p$
ratio can be obtained from fits to the restricted
average dE/dx distributions, such as those shown in Fig.~2a and b
for the rigidity interval from 4.5 -- 6 GV.

The results are summarized in Table \ref{pbar_results}.
In order to obtain the number of protons and antiprotons at
the top of the atmosphere, we correct for particle production
in the atmosphere (total average column density 7.2 g/cm$^2$) above
the instrument. In addition, corrections for interaction and
annihilation losses of protons and antiprotons in the atmosphere
and in the instrument (maximum column density 5.7 g/cm$^2$ above the
lower ToF counter for a
vertically traversing particle) are applied.
The corrections assume that all particles that interact inside the gondola
are rejected by our selection criteria.

The calculated background of secondary antiprotons and protons
produced in the atmosphere was based on
Pfeifer et al.\ \cite{pfeifer96} for the antiprotons and
Papini et al.\ \cite{sec_proton_ref} for the protons. Interaction
and annihilation losses
are based on the measured cross sections quoted in
Kuzichev et al.\ \cite{kuzichev94} and
Denisov et al.\ \cite{Denisov73}, accounting in detail
for the total material traversed by a particle in passing through
the atmosphere, aluminum shell, and detector material.
The number of antiprotons and protons in each energy bin obtained
after applying all of these corrections are shown in Table
\ref{pbar_results}, along with the
resulting $\overline{p}/p$ ratios. The errors
quoted in this table are purely statistical. Systematic
errors resulting from uncertainties in correcting the particle numbers
in the instrument to the top of the atmosphere and in the background
due to particle mis-identification are estimated to be less than 4\%
of the $\overline{p}/p$ ratio.

\begin{table} [b]
\caption{Event selection results and $\overline{p}/p$ ratios (in
10$^{-4}$). R is the measured rigidity at the spectrometer and T is
the corresponding kinetic particle energy at the top of
the atmosphere. N$_p$ and N$_{\overline{p}}$ are the number of observed
protons and antiprotons for each energy bin, respectively.
 N$_p^{corr.}$ and N$_{\overline{p}}^{corr.}$ are the extrapolated number
 of protons and antiprotons at the top of the atmosphere. The
pion/muon background due to tails in
the $\langle dE/dx_{res} \rangle$ distributions
in the five rigidity bands is (0.2, 0.3, 0.4, 0.7, 0.8) counts
and is included in the corrections. 
}
\begin{ruledtabular}
\begin{tabular}{ccrrrrr}
\multicolumn{1}{c}{R (GV)}
& \multicolumn{1}{c} {T (GeV)}
& \multicolumn{1}{c} {N$_p$}
& \multicolumn{1}{c} {N$_{\overline{p}}$}
& \multicolumn{1}{c} {N$_p^{corr.}$}
& \multicolumn{1}{c} {N$_{\overline{p}}^{corr.}$}
& \multicolumn{1}{c} {$\overline{p}/p$ ratio}\\\hline
\\
4.5 -- 6.0  & 3.7 -- 5.1 &119361 &18 &124814& 13.9 &1.11$^{+0.50}_{-0.39}$\\
6.0 -- 10.0 & 5.1 -- 9.1 &141447 &23 &148952& 16.9 &1.13$^{+0.46}_{-0.37}$\\
10.0 -- 15.0&9.1 -- 14.1 & 60727 &21 &64971 & 18.9 &2.91$^{+1.01}_{-0.81}$\\
15.0 -- 25.0&14.1 -- 24.1& 37742 &15 &40141 & 12.9 &3.21$^{+1.42}_{-1.10}$\\
25.0 -- 50.0&24.1 -- 49.1& 8773  &1  &9090  & 0    &$<2.1 (90\%)$\\
\end{tabular}
\end{ruledtabular}
\label{pbar_results}
\end{table}

\begin{figure}[th]
\includegraphics[width=0.999\linewidth]{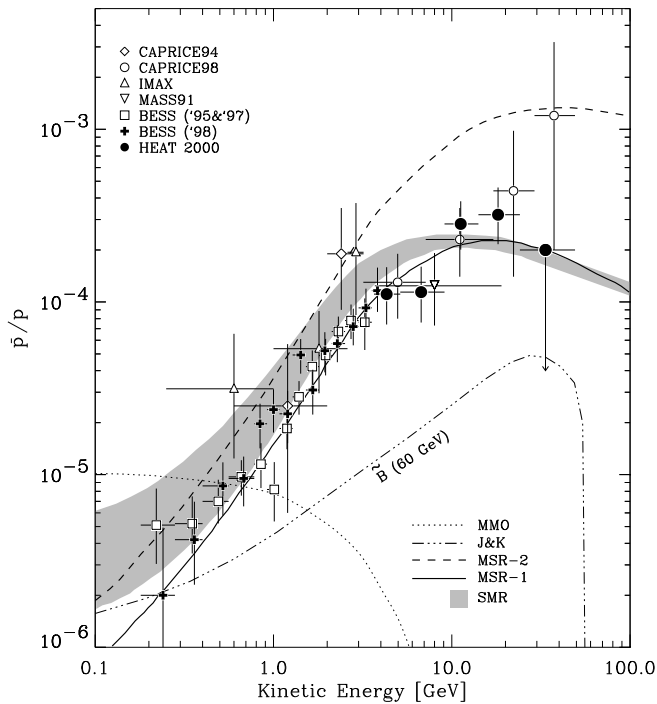}
\caption{Compilation of observed $\overline{p}/p$ flux ratios at
the top of the atmosphere, compared with model calculations
for secondary and primary antiproton production:
BESS 95\&97 \cite{bess00},
BESS \cite{bess_pp},
IMAX \cite{imax},
MASS91 \cite{mass2},
CAPRICE94 \cite{caprice94},
CAPRICE98 \cite{caprice98}.
The calculations of the $\overline{p}/p$ ratio are from 
\cite{msr98} (MSR-1, MSR-2) and \cite{simon}(SMR). Possible primary
contributions to the $\overline{p}/p$ spectrum arising from
evaporating primordial black holes \cite{maki96} (MMO) and from
neutralino annihilation \cite{jk} (J\&K) are also shown.
}
\label{pbarp}
\end{figure}

Our results are shown in Fig. \ref{pbarp}, along with previous
measurements by others, and a number of
recent calculations of the $\overline{p}/p$ ratio resulting from
secondary production of antiprotons in the interstellar medium.
Only recent measurements have been included in this figure \cite{bess00,
bess_pp,imax,mass2,caprice94,caprice98}.

Many predictions for the $\overline{p}/p$ ratio have been
published over the years. We show here theoretical curves
that are consistent with the now well measured flux ratio
in the low energy region around 1 GeV. The result of
calculations by Simon et al. \cite{simon} are shown
in the figure as a shaded band. The calculations are based
on the leaky box model and the uncertainties in the flux prediction,
reflected by the band in the figure, are primarily uncertainties
in the galactic path length distribution.
The dashed and solid line in Fig \ref{pbarp} show the results
of calculations by Moskalenko et al. \cite{msr98} within a
self-consistent CR propagation model. The dashed line represents
the case of a proton injection spectrum that is much harder than
locally observed, which has been proposed to explain the observed
high continuum gamma-ray emission above $\sim$ 1 GeV \cite{hunter97}.
A standard proton injection spectrum, consistent with the locally
observed one, is reflected in the solid line.
The sensitivity of the  $\overline{p}/p$ ratio to the nucleon
injection spectrum above a few GeV makes antiproton measurements
at energies above a few GeV an important test for CR models.
Our data are in good agreement with the `standard spectrum'
calculations \cite{msr98} at high energy, and do not support an
antiproton to proton ratio approaching 10$^{-3}$ at energies
above 20 GeV, in contrast to recent CAPRICE measurements \cite{caprice98}.
Our result does not support models which are based on hard
nucleon injection spectra. At energies covered by the
measurements presented here, secondary $\overline{p}$ production
with a nucleon injection spectrum consistent with the locally
observed one describes the data well.

The HEAT-pbar instrument is scheduled for additional balloon flights
and we expect to statistically improve the data and to further clarify
the experimental situation.

\begin{acknowledgments}
This work was supported by NASA grants No. NAG 5-5058,
No. NAG 5-5220, No. NAG 5-5223, and No. NAG 5-5230, and
by financial assistance from our universities.
We wish to thank the National Scientific Balloon Facility and the NSBF
launch crews for their excellent support of balloon missions and
we acknowledge contributions from D. Kouba, M. Gebhard, S. Ahmed,
and P. Allison.
\end{acknowledgments}

\end{document}